%% file: heavyq94.tex
\documentstyle[twoside,fleqn,espcrc2,epsf]{article}

\newcommand{\ewxy}[2]{\setlength{\epsfxsize}{#2}\epsfbox[10 60 640 570]{#1}}

\newcommand{\be}{\begin{equation}}
\newcommand{\ee}{\end{equation}}
\newcommand{\order}{{\cal O}}

\newcommand{\pv}{{\bf p}}
\newcommand{\fig}[1]{Fig.~\ref{#1}}

\newcommand{\eq}[1]{(\ref{#1})}

\newcommand{\Bv}{{\bf B}}
\newcommand{\Ev}{{\bf E}}

\newcommand{\delv}{{\bf \del}}
\newcommand{\sigmav}{{\mbox{\boldmath$\sigma$}}}
\newcommand{\msb}{{\overline{\rm MS}}}
\newcommand{\nl}{\nonumber \\}
\newcommand{\delfour}{{\Delta^{(4)}}}

\newcommand{\Ups}{\Upsilon}

\newcommand{\del}{{\bf \Delta}}
\newcommand{\delsq}{\Delta^{(2)}}

\newcommand{\mev}{{\rm MeV }}
\newcommand{\gev}{{\rm GeV }}

\newcommand{\ainv}{a^{-1}}

\newcommand{\oneS}{1{\rm S}}
\newcommand{\twoS}{2{\rm S}}

\newcommand{\oneP}{1{\rm P}}
\newcommand{\twoP}{2{\rm P}}

\newcommand{\coeff}[2]{{{#1}\over {#2}}}

\newcommand{\picplus}[2]{
\put(#1,#2){\line( 0, 1){1}}
\put(#1,#2){\line( 0,-1){1}}
\put(#1,#2){\line(-1, 0){1}}
\put(#1,#2){\line( 1, 0){1}}
}

\newcommand{\SW}{SW }

\newcommand{\HW}{IHW }
\newcommand{\KT}{KT }
\newcommand{\FNAL}{Fermilab }
\newcommand{\CDHW}{CDHW }
\newcommand{\boks}{Bodwin, Kim and Sinclair}

\newcommand{\gape}{$\raisebox{-.6ex}{$\stackrel{\textstyle{>}}{\sim}$}$}
\newcommand{\lape}{$\raisebox{-.6ex}{$\stackrel{\textstyle{<}}{\sim}$}$}
\newcommand{\nf}{n_f}
\newcommand{\amsbar}{{\alpha_\msb^{(5)}(M_Z)}}
\newcommand{\mbare}{{M_0}}
\newcommand{\ccb}{{c\bar c}}
\newcommand{\bbb}{{b\bar b}}
\newcommand{\lqcd}{{\Lambda_{QCD}}}
\newcommand{\alv}{{a\Lambda_V}}

\newcommand{\AmS}{{\protect\the\textfont2
  A\kern-.1667em\lower.5ex\hbox{M}\kern-.125emS}}

\title{Latest Results from Heavy Quark Simulations}

\author
{
 John H. Sloan
  \address
   {
    SCRI, 
    Florida State University,
    Tallahassee, Fl 32306-4052,
    USA
   }
}

\begin{document}

\begin{abstract}
I review the status of $\bbb$ and $\ccb$ calculations, numerical and 
analytic.  I discuss the
extraction of $\alpha_s$ and quark masses from spectrum calculations.
I compare the NRQCD and Improved Heavy Wilson 
formulations of heavy quarks, and discuss recent calculations 
using a Heavy Staggered formulation.
\end{abstract}

\maketitle

\section{Introduction}

I have taken my charge in this review to be two-fold: to review
this year's results from heavy-heavy meson calculations and to compare
the two main formalisms for treating heavy quarks, NRQCD\cite{cornell-action}
and the mass dependent improvement program of the Fermilab 
group\cite{fnal-action} 
(which I call Improved Heavy Wilson(IHW) fermions).  
I will try to minimize my overlap with previous 
reviews\cite{lepage-lat91,mackenzie-lat92,davies-lat93}.
The messages that I hope people take from the theory part of this talk 
are that \HW fermions {\it are} (improved) Wilson fermions
(with an on-shell renormalization conditions), that
NRQCD and \HW differ only by negligible irrelevant 
operators at large $am$,
and that NRQCD actions have an extra symmetry (the heavy quark 
symmetry) which makes them much less messy to deal with but causes
them to blow up at small $am$.  

From the new results, my message is
that we probably understand what we're doing in onium systems; 
discretization errors, quenching effects, spin splittings and
perturbation theory all seem to behave as expected.  If this is true, 
then we are probably
within one to three years of being able present an extensively cross-checked
determination of $\amsbar$, with uncertainty at the 1-2\% level and all
ingredients calculated in several different ways, to the 
rest of the particle physics community.
Another message is that tadpole-improved coefficients\cite{lmac} are necessary
for perturbative improvement to work well.
Finally, there is a discrepancy between typical lattice spacings 
determined from $\bbb$
physics and those from light spectroscopy (both quenched and dynamical);
we should be aware of this as a potential source of systematic error
in light hadron calculations.

The reader should assume that any result reviewed here is preliminary, 
unless it appears elsewhere.

\section{\HW and NRQCD: Compare, Contrast}

\subsection{A Toy Model}

The most important feature of heavy quark ($am \gape 1$) 
calculations on the lattice is that, for a lattice O(4) invariant 
action, boost invariance of on-shell correlation functions is strongly 
broken, even for soft boosts of particles with (spatial) momentum 
much smaller than the cutoff.  This is because all components of the 
on-shell four-momentum cannot be made small simultaneously; the energy
will always be of order the cutoff or higher.  To restore low-momentum
boost invariance without taking the cutoff much larger than the mass, 
(lattice) boost invariance must be broken in the bare action; 
lattice O(3) is the appropriate symmetry group for heavy quark bare
actions.
NRQCD implements the breaking O(4)$\rightarrow$O(3) directly;
NRQCD is a non-relativistic effective action.
In the \HW formalism the breaking is more subtle; the \HW formalism can
be thought of as just the normal improvement program for Wilson fermions,
except the renormalization conditions are imposed with the heavy quark
{\it on-shell}\cite{kronfeld-lat92}.  The Lorentz breaking arises 
because the four-momentum of a heavy quark is very anisotropic;
a `preferred frame' is chosen which is just the rest frame of the heavy
quark.

To illustrate this, consider the dispersion relation of an on-shell (free) 
naive fermion;
\be
(am)^2 + \sum_\mu \sin^2(ap_\mu) = 0.
\label{free_nv_disp}
\ee
If $am \ll 1$, all of the components of the dimensionless momentum can be
chosen small, so that $\sin$ can be replaced with its
expansion;
\be
(am)^2 + (ap)^2 - \coeff{1}{3}\sum_\mu (ap_\mu)^4 + \dots  = 0,
\label{free_nv_disp_smallm}
\ee
where $(ap)^2 $=$ \sum_\mu (ap_\mu)^2$.  The first two terms in 
\eq{free_nv_disp_smallm} represent the continuum dispersion relation, while
the third and later terms are cutoff effects.  Since these are much less
than one, \eq{free_nv_disp_smallm} is approximately 
Lorentz invariant at low momenta.

When $am$ is large, however, this fails.  At least one component 
(the energy) of the 4-momentum in \eq{free_nv_disp} must be large, so as 
to cancel the mass term, and the $\sin$ of this component cannot be 
expanded.  This leads to the dispersion relation
\be
(am)^2 + \sin^2(ap_0) + (a\pv)^2 - \coeff{1}{3}\sum_i (a\pv_i)^4 + \dots  = 0,
\label{free_nv_disp_bigm}
\ee
where $\pv$ is the spatial momentum.  Even at low momenta, 
\eq{free_nv_disp_bigm} is not Lorentz invariant, and so the 
bare action that leads to it cannot be used directly to
calculate S-matrix amplitudes.  Discretization errors
in the temporal and spatial directions are different; to obtain
the correct behavior near the mass-shell
the temporal and spatial terms in the action must be adjusted separately.
This results in an action which only has lattice O(3) symmetry, but
which has O(4) invariant correlation functions at low momentum.

The dispersion relation \eq{free_nv_disp_bigm} has
behavior very similar that of \HW 
fermions\cite{kronfeld-lat92,mackenzie-lat92}; the 
static mass, $M_1$, depends logarithmically on $am$ when $am \gg 1$, while the 
kinetic mass, $M_2$, goes linearly with $am$.  A Lorentz invariant theory
has $M_1$=$M_2$=$M_3$=$\dots$, where $M_3$ is the mass appearing in the $\pv^4$
term in the relativistic low momentum dispersion relation. $M_1$=$M_2$ can 
be recovered
by rescaling the spatial derivatives relative to the temporal derivatives;
this is equivalent to the `asymmetric $\kappa$' prescription in 
\cite{kronfeld-lat92}.  $M_3$ must be adjusted by adding a $\pv^4$ improvement
term to the action; note that this is a lattice O(3), rather than lattice
O(4), invariant operator.  For Wilson fermions, the clover term,
$\bar\psi \sigma_{\mu\nu} F^{\mu\nu} \psi$, is the lowest
order improvement operator needed in the action\cite{sw-action}.  When 
$am \gg 1$, the
asymmetric physical renormalization point requires that $\sigma\cdot F$ be 
broken up into a sum of electric and magnetic  O(3) invariant 
representations, with separately tuned coefficients for the electric and 
magnetic terms.  I will call this the "asymmetric clover" level of 
improvement for the \HW formulation.

The \FNAL group has pointed out that only a modified
Lorentz invariance is needed when studying processes involving valence
heavy quarks; rather than using an asymmetric $\kappa$ to tune $M_1$=$M_2$,
one can shift the zero of energy in the continuum theory by $M_2-M_1$ for 
each quark without changing the physics.  
If this
is true, then $\kappa$ can be kept symmetric as long as $M_1$ is ignored.
This is typically done both in \HW and NRQCD.  Asymmetric improvement terms
are still needed at higher orders; for example 
an asymmetric clover
coefficient is needed to get both spin-spin and spin-orbit interactions right.


\subsection{Similarities and Differences}

\HW and NRQCD are very similar formalisms for simulating heavy quarks.
Both break lattice O(4) invariance to lattice O(3) in the bare action to
recover boost invariant low momentum correlation functions.  Both rely
heavily on an improved action program to eliminate discretization errors,
and the systematic errors in both can be expanded in $\lqcd/M_Q$ 
(where $\lqcd\sim 500\mev$).  Although the \HW is formulated
in terms of Dirac spinors, it can be thought of in the
Foldy-Wouthuysen-Tani(FWT) transformed basis of NRQCD.  

There is an intrinsic difference between the two formalisms, however.  
In the FWT basis,  NRQCD throws away the 
coupling between large and small components of the 4-component spinors,
resulting in a block diagonal action involving (2 sets of) 2-component 
Pauli spinors.  This imposes an additional symmetry on the action (the
heavy quark symmetry), and causes the NRQCD action to be non-renormalizable
when $am\rightarrow 0$.  The advantage of this procedure is that $m$ can
be removed as a scale before discretization.  The \HW action does
not eliminate the couplings between large and small components and so
is renormalizable (going over into an improved Wilson action as 
$am\rightarrow 0$), at a cost of having to treat discretization errors
to all orders in $am$.

To illustrate the difference, start with the free continuum Dirac action
(note that neither NRQCD nor \HW is actually derived as below,
but the basic principles are the same);
\be
S_{Dirac} = \bar\psi
\left(
\begin{array}{cc}
\partial_t + m & i\sigmav\cdot\pv    \\
i\sigmav\cdot\pv &  -\partial_t + m 
\end{array}
\right)
\psi.
\label{dirac_action}
\ee
A continuum FWT transformation can now be performed to some order in
$v^2 $=$ (\pv/m)^2$, resulting in (for the lowest order transformation)
\be
\bar\psi
\left(
\begin{array}{c}
\partial_t + m + \coeff{\pv^2}{2m} + \dots\quad \dots    \\
\dots\quad -\partial_t + m + \coeff{\pv^2}{2m} + \dots
\end{array}
\right)
\psi,
\label{FWT_action}
\ee
where $\dots$ represents terms which lead to higher order corrections
to the dispersion relation, such as $\pv^4/m^3$.
If we discretize \eq{FWT_action} directly, we obtain
\be
\bar\psi
\left(
\begin{array}{ll}
\ainv\sinh(aE) + m + \coeff{\pv^2}{2m} + \dots\quad \dots    \\
\dots\quad -\ainv\sinh(aE) + m + \coeff{\pv^2}{2m} + \dots
\end{array}
\right)
\psi,
\label{HW_action}
\ee
This is the \HW approach; as m gets large, $\sinh(aE)$ and hence $aE$ must 
also become
large, leading to large discretization errors.  In the NRQCD approach,
the higher order terms ($\dots$) in \eq{FWT_action} are set explicitly 
to zero.  This makes \eq{FWT_action} block-diagonal; the transformation
$\bar\psi \rightarrow \bar\psi e^{m\gamma_0t}$, 
$\psi \rightarrow e^{-m\gamma_0t}\psi$ can now be performed on 
\eq{FWT_action}, 
resulting in an action
\be
S_{NR} =
\bar\psi
\left(
\begin{array}{ll}
\partial_t + \coeff{\pv^2}{2m} \quad 0  \\
0 \quad -\partial_t + \coeff{\pv^2}{2m} 
\end{array}
\right)
\psi.
\label{NR_action}
\ee
This is the NRQCD approach; $\coeff{a\pv^2}{2m}$ is \order($mv^2$), so
there are no problems discretizing \eq{NR_action}.  
When $am\rightarrow 0$, the symmetry imposed by the zero entries in
\eq{NR_action} is no longer valid; the theory cannot be renormalized.
This method for removing the zero of energy doesn't work
for IHW; $\gamma_0$ doesn't commute with the off-diagonal components
of \eq{FWT_action}.

\subsection{Pros and Cons}

In IHW, discretization errors must be treated to all orders in $am$.  
In NRQCD, $m$ has been removed as a dynamical energy scale, and discretization
errors can be expanded in $a\pv$.  This means that:

\begin{itemize}

\item{
In IHW, the tree-level coefficients of terms in the 
action are complicated, non-intuitive functions of $am$.  In NRQCD,
these coefficients are simple functions, for which non-relativistic
intuition can be easily applied.
}

\item{
NRQCD is a non-renormalizable theory; the coefficients in the NRQCD
action blow up when $am\lape .6$ (see \fig{fig:morning}).
An \HW action is renormalizable;
as $am\rightarrow 0$ it goes smoothly over
to a perturbatively improved Wilson action.
}

\item{
Because NRQCD blows up at small $am$, charm can only be simulated at
$\beta\lape 5.85$.
\HW simulations can be performed at any $\beta$; brute force can be used
to reduce discretization errors and there is a much larger range of $\beta$
for which charm and bottom quarks can be simultaneously studied.
}

\item{
Because of the complicated coefficients, the \HW action is harder
to derive than NRQCD.  To date, only the terms which contribute at
leading order in $v^2$ to the spin-averaged and spin dependent spectra 
have been published for the tree-level \HW action.  The 
relativistic corrections to these terms have been published, also 
at tree level, for NRQCD.
}

\item{
\HW fermions cost more CPU time than NRQCD;  \HW propagators 
solve an elliptic partial differential equation(PDE), while
NRQCD solves a parabolic PDE.
I suspect the \HW inverter can be accelerated.
When $am$ is large, the \HW PDE is close to parabolic;
there should be a way to use this.
}

\end{itemize}

\subsection{Levels of Improvement}

I find it easiest to use the NRQCD action
derived in \cite{cornell-action} to catalogue improvement terms
in the NRQCD or \HW actions and to understand their
effects.  The \order($mv^4$) NRQCD action is given by
\begin{eqnarray}
 S &=& \bar\psi (\Delta_t + H) \psi. \nl
 H &=& - {\delsq\over2\mbare}
\nl & &
 - C_1 \frac{(\delsq)^2}{8(\mbare)^3}
 + C_2 \frac{ig \left(\delv\cdot\Ev - \Ev\cdot\delv\right)}{8(\mbare)^2}
\nl & &
 - C_3 \frac{g \sigmav\cdot(\delv\times\Ev - \Ev\times\delv)}{8(\mbare)^2}
 - C_4 \frac{g\sigmav\cdot\Bv}{2\mbare}\,
\nl & &
 + C_5 \frac{a^2\delfour}{24\mbare}
 - C_6 \frac{a(\delsq)^2}{16n(\mbare)^2},
\label{nrqcd_action}
\end{eqnarray}
where $\delsq$ is the discrete laplacian, and $\delfour$=$\sum_i\Delta_i^4$.
For HH mesons, the $C_i$ terms all contribute \order($mv^4$) to the energy; 
the kinetic energy
term is, of course, \order($mv^2$).  Further improvement operators are
suppressed by more powers of $v^2 \sim \lqcd/M$, which is about $.1$ for
$\bbb$ and $.3$ for $\ccb$.  
The counting for improvement operators is
different in HL mesons; see \cite{cornell-action} for details.
All operators are tadpole-improved; $C_i$=$1$ at tree-level.

The terms in \eq{nrqcd_action} are easily identified from low energy
electrodynamics; $C_1$ multiplies the leading relativistic correction
to the dispersion relation, $C_2$ the Darwin term, $C_3$ the spin-orbit
coupling (causing fine-structure), and the $C_4$ term gives rise to the 
spin-spin interaction (causing hyperfine splittings).  The $C_5$ and $C_6$
terms correct the leading discretization errors in the spatial and temporal
directions, respectively.

As more accuracy and physical information is required from a 
heavy quark calculation, more improvement terms can be turned on in 
the action.
For every NRQCD action, there is a corresponding \HW action\cite{fnal-action} 
in which
similar improvement terms have been turned on.  At large $am$,
the effects of improvement terms which have not yet been turned on
should be the same size in \HW and NRQCD.  At small $am$, however,
the \HW action recovers lattice O(4) invariance, so the effects of
lattice O(4) breaking improvements are small.  In NRQCD language,
the Wilson action gets the term corresponding to $C_1$ right at
small $am$, without needing improvement.  In this regime,
it is much harder to estimate
the \HW systematic errors and compare to \order($mv^4$) NRQCD, for example.  
El-Khadra has used potential models to estimate these
errors in the \FNAL group's calculations\cite{khadra-lat94}, but 
it is also important
to obtain simulation results from the \HW action equivalent to 
\eq{nrqcd_action}, so that the two methods can be directly compared.

The simplest action to calculate with is just \order($mv^2$) NRQCD, in which
all the $C_i$ terms in \eq{nrqcd_action} are turned off.  This 
is a spin-symmetric action;
spin-averaged splittings will have errors at the $v^2$ level, but
spin dependent splittings will be zero.  The equivalent \HW action
is just the normal Wilson action; spin splittings are non-zero but
much too small.  Note that the $C_1$ and $C_2$ terms contribute to
the kinetic mass at the $v^2$ level; if more precision is
needed for $M_2$ then these improvements (and probably $C_5$ and $C_6$)
should be turned on.

To study spin splittings,  the $C_3$ and $C_4$ improvements
must be turned on.  In IHW, this is equivalent to turning 
on the clover term; the magnetic part corresponds to the
$C_4$ term, while the electric part is the sum of $C_2$ and $C_3$
improvements.  The Sheikholeslami-Wohlert action\cite{sw-action}
has the correct $C_4$ improvement, but $C_3$
and $C_2$ are somewhat wrong.  This is fixed by going to the
asymmetric clover action; at tree-level, the $C_2$, $C_3$, and $C_4$ terms 
are all correctly included.

The most complicated action in use is the $mv^4$ NRQCD action in
\eq{nrqcd_action}.  Errors in spin averaged splittings and 
kinetic masses should be at the $v^4$ level, while spin splittings
still have $v^2$ errors.  The corresponding \HW action is not
published; it is important that \HW calculations be done at this order 
to compare with NRQCD results.

If one wants to reduce errors in the spin splittings to \order($v^4$), 
then one
must go to the level of improvement in \cite{cornell-action}.
This is the first level at which one-loop results are needed
for the $C_i$.  Since $\alpha_s $ should be about the same size
as $v^2$, the $\alpha_s mv^4$ errors arising from the use of tree-level
$C_i$ should be about the same size as $mv^6$ improvement terms.
(Note that I have used this to
simplify the discussion of errors above. $v^4$ errors really means
$v^4$ and $\alpha_s v^2$  errors; both should be the same size.)  
Until the one-loop corrections to $C_3$ and $C_4$ (or their \HW 
equivalents) are known, 
this level of improvement cannot be fully implemented.  

The results presented this year have different levels of improvement
for different groups.  
The $mv^2$ NRQCD action was used by the UK(NR)QCD group\cite{uknrqcd}, and by
\boks\cite{sinclair-lat934}, who have performed a 
very nice calculation of matrix elements involved in $\bbb$ decay.
The \KT group\cite{kt-alphas,onogi-lat93} and Wingate 
{\it et al.}\cite{cdhw} (referred to as \KT and \CDHW below)
both use a Wilson action, which should have similar
errors.  The \FNAL group\cite{khadra-lat94} uses the 
SW action; the NRQCD collaboration 
\cite{davies-lat94,nrqcd-alphas,nrqcd-mb} uses the $mv^4$ NRQCD action.
Both use tadpole improvement, which is essential.

\section{Perturbation Theory}

Perturbative calculations are an essential ingredient in both the
NRQCD and \HW formalisms.
To date, NRQCD and \HW simulations have been done using only tree-level
coefficients for the improvement terms in the action.
The next level of relativistic corrections,
\order($mv^6$), however, are about the same size as the one-loop corrections
to the current action, \order($\alpha_s mv^4$), so loop calculations are 
needed to improve the actions beyond \order($mv^4$).  
Besides the action, perturbation theory is needed for non-spectroscopic
results.  For example, a calculation of the pole or $\msb$ quark masses
requires a perturbative calculation of the quark dispersion relation,
while wave function and current renormalizations will be needed for 
quantities such as $f_B$ or the leptonic width of onium.

The \FNAL group has an ongoing program to perform 
perturbative calculations for the \HW 
formalism\cite{khadra-lat94,kronfeld-lat93,kronfeld-lat94,fnal-loop},
while Morningstar has performed most of the recent perturbative calculations
in NRQCD\cite{morning-notad,morning-qstar}.  
Both calculate the dispersion relation of the heavy quark; 
matching to the low energy continuum dispersion relation then fixes the
$C_1$, $C_5$, and $C_6$ improvements, as well as the corrections
needed to determine the pole or $\msb$ mass\cite{nrqcd-mb}.
The main new results this year are the inclusion of the Lepage-Mackenzie
$q^*$ in these perturbative 
calculations\cite{morning-qstar,khadra-lat94,kronfeld-lat94}.


\begin{figure}[htb]
{\setlength{\epsfxsize}{60mm}\epsfbox[30 440 440 750 ]{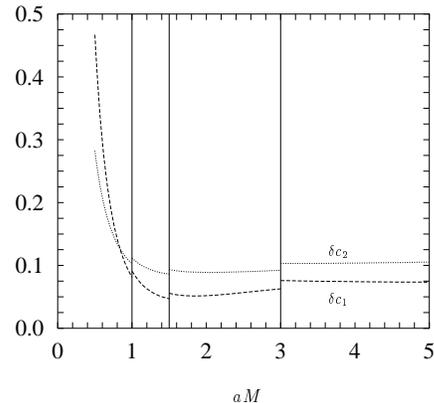}}
\caption{Calculation by Morningstar
{\protect\cite{morning-qstar}} 
of the 1-loop perturbative correction 
to the coefficients $C_1$, $C_5$, and $C_6$
in the NRQCD action as a function of aM.  
Note that Morningstar's $c_1$ corresponds to a combination of $C_1$
and $C_6$, while $c_2$ is $C_5$ in {\protect\eq{nrqcd_action}}.}
\label{fig:morning}
\end{figure}


The corrections to the action appear well behaved in both formalisms; 
Morningstar's results for the one loop contribution to the $C$'s are shown 
in \fig{fig:morning}.  The one-loop
correction is roughly 10\% for $am>.8$;
this is a good indication that perturbation theory is working.  
Notice that the corrections begin to grow rapidly
below $am<.6$; this is where NRQCD begins to ``blow up.''


\begin{figure}[htb]
{\setlength{\epsfxsize}{60mm}\epsfbox[30 120 550 484 ]{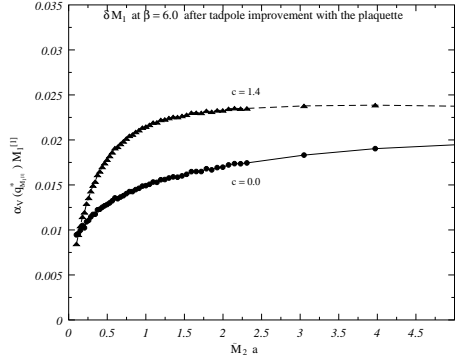}}
\caption{Perturbative correction to the static mass ($M_1$) for 
both Wilson ($c$=$0$) and \SW ($c$=$1.4$) actions as a function of
the kinetic mass ($M_2$){\protect\cite{mertens-priv}}.
 }
\label{fig:mertens}
\end{figure}


The \FNAL group has similar results for the perturbative correction to
the static mass; their results for both \SW and Wilson actions are
shown in \fig{fig:mertens}.  Again, the result is very 
well-behaved.  This calculation can be combined with simulation results
to determine the quark's pole mass\cite{khadra-lat94}.


\section{Heavy Staggered Fermions}

This year, the \KT group has introduced a third heavy fermion formalism;
Heavy Staggered (HS) fermions\cite{aoki-lat94}.  They have performed 
simulations at both quenched $\beta$=$6.0$ and 
at $n_f$=$2$ staggered, $\beta$=$5.7$, $am_s$=$.01$\cite{kt-ens}, with
bare valence masses ranging from $am_Q$=$.2$ to $am_Q$=$.5$.  They used 
static, rather than kinetic, 
masses to match with experiment, but at these values of the $am$
this is probably not a huge effect.
As with most staggered fermion calculations, they had to deal with a
daunting number of meson creation operators; they used 128 operators 
grouped into 36 irreducible representations!


\begin{figure}[htb]
{\setlength{\epsfxsize}{60mm}\epsfbox[58 181 472 422 ]{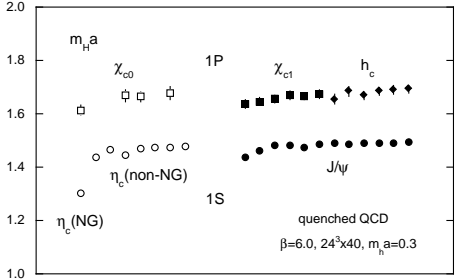}}
\caption{Heavy staggered $\ccb$ spectrum.  NG refers to flavor singlet 
pseudoscalar.  Different points correspond to different irreducible
lattice representations of the continuum state{\protect\cite{aoki-lat94}}.
}
\label{fig:kt-spect}
\end{figure}


Results for the $\ccb$ spectrum (averaged within irreducible representations)
are shown in \fig{fig:kt-spect}.  As can be seen in the figure, the masses of
the states are dependent on the particular operator used.  Furthermore,
this operator dependence seems to go like the square of the heavy 
mass\cite{aoki-lat94}.  The most pronounced of these splittings is between
the flavor singlet and flavor non-singlet versions of the $\eta_c$; the
\KT group quotes separate results for the `NG' and `non-NG' pseudoscalars.


\begin{figure}[htb]
{\setlength{\epsfxsize}{60mm}\epsfbox[48 184 467 460 ]{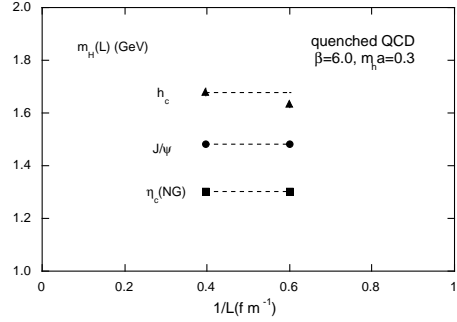}}
\caption{Finite volume dependence of HS $\ccb$ states.
Note the large shift in the $h_c$ state. {\protect\cite{aoki-lat94}}
}
\label{fig:kt-volume}
\end{figure}


A result that may be of concern to those using other formulations for $\ccb$
is shown in \fig{fig:kt-volume}, which compares HS results at quenched 
$\beta$=$6.0$ on $16^3$x$40$ and $24^3$x$40$ lattices.  The change of about 
3\% in the mass of the $h_c$ results in a 20\% change in the $\oneS$-$\oneP$
splitting.  The smaller lattice is roughly the same physical size as
most of those which have been used for $\ccb$ simulations;
an explicit check of volume dependence in 
$\ccb$ spectroscopy should probably be done using the \HW or NRQCD 
formulation.  This is much less likely to be a problem in $\bbb$ 
spectroscopy; the $\bbb$ mesons are much smaller than $\ccb$.

The \KT group has extracted a value of
$\amsbar$ from the $\ccb$ $\oneP$-$\oneS$ splitting; they obtain 
$.111(4)(2)$ from their quenched results and an upper bound
of $.119(3)(3)$ at $n_f$=$2$,
where the first uncertainty is from the scale determination and $n_f$
correction and the second comes from  flavor breaking.
They have asked me to stress that, because of
the strong finite volume dependence of the $h_c$ mass, the $n_f$=$2$
result should only be considered an upper bound.

\section{Onium Spectroscopy}

\subsection{$\bbb$}

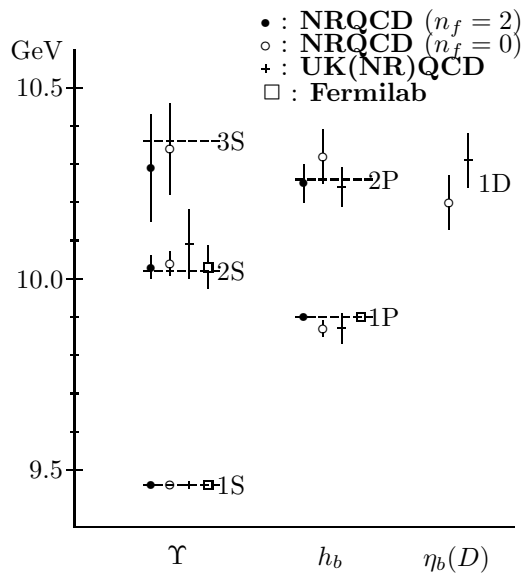
\begin{figure}[htb]
\begin{center}
\setlength{\unitlength}{.020in}
\begin{picture}(115,140)(10,930)
\put(15,935){\line(0,1){125}}
\multiput(13,950)(0,50){3}{\line(1,0){4}}
\multiput(14,950)(0,10){10}{\line(1,0){2}}
\put(12,950){\makebox(0,0)[r]{9.5}}
\put(12,1000){\makebox(0,0)[r]{10.0}}
\put(12,1050){\makebox(0,0)[r]{10.5}}
\put(12,1060){\makebox(0,0)[r]{GeV}}
\put(15,935){\line(1,0){115}}


\put(63,1049){\makebox(0,0)[l]{$\,\Box$ : {\bf Fermilab}}}
\picplus{65}{1055}
\put(69,1055){\makebox(0,0)[l]{: {\bf UK(NR)QCD}}}
\put(65,1061){\makebox(0,0)[tl]{\circle{2}}}
\put(69,1061){\makebox(0,0)[l]{: {\bf NRQCD $(n_f = 0)$ }}}
\put(65,1067){\makebox(0,0)[tl]{\circle*{2}}}
\put(69,1067){\makebox(0,0)[l]{: {\bf NRQCD $(n_f = 2)$ }}}

\put(42,930){\makebox(0,0)[t]{$\Upsilon$}}
\put(56,946){\makebox(0,0){1S}}
\multiput(33,946)(3,0){7}{\line(1,0){2}}
\put(35,946){\circle*{2}}
\put(40,946){\circle{2}}
\picplus{45}{946}
\put(49,945){\framebox(2,2)}

\put(56,1002){\makebox(0,0){2S}}
\multiput(33,1002)(3,0){7}{\line(1,0){2}}
\put(35,1003){\circle*{2}}
\put(35,1003){\line(0,1){3}}
\put(35,1003){\line(0,-1){3}}
\put(40,1004){\circle{2}}
\put(40,1005){\line(0,1){2}}
\put(40,1003){\line(0,-1){2}}
\picplus{45}{1009}
\put(45,1010){\line(0,1){8}}
\put(45,1008){\line(0,-1){8}}
\put(49,1002){\framebox(2,2)}
\put(50,1004){\line(0,1){4.6}}
\put(50,1002){\line(0,-1){4.6}}

\put(56,1036){\makebox(0,0){3S}}
\multiput(33,1036)(3,0){7}{\line(1,0){2}}
\put(35,1029){\circle*{2}}
\put(35,1029){\line(0,1){14}}
\put(35,1029){\line(0,-1){14}}
\put(40,1034){\circle{2}}
\put(40,1035){\line(0,1){11}}
\put(40,1033){\line(0,-1){11}}

\put(82,930){\makebox(0,0)[t]{$h_b$}}

\put(96,990){\makebox(0,0){1P}}
\multiput(73,990)(3,0){7}{\line(1,0){2}}
\put(75,990){\circle*{2}}
\put(80,987){\circle{2}}
\put(80,988){\line(0,1){1}}
\put(80,986){\line(0,-1){1}}
\picplus{85}{987}
\put(85,988){\line(0,1){3}}
\put(85,986){\line(0,-1){3}}
\put(89,989){\framebox(2,2)}

\put(96,1026){\makebox(0,0){2P}}
\multiput(73,1026)(3,0){7}{\line(1,0){2}}
\put(75,1025){\circle*{2}}
\put(75,1025){\line(0,1){5}}
\put(75,1025){\line(0,-1){5}}
\put(80,1032){\circle{2}}
\put(80,1033){\line(0,1){6}}
\put(80,1031){\line(0,-1){6}}
\picplus{85}{1024}
\put(85,1025){\line(0,1){4}}
\put(85,1023){\line(0,-1){4}}

\put(125,1025){\makebox(0,0){1D}}
\put(115,930){\makebox(0,0)[t]{$\eta_b(D)$}}
\put(113,1020){\circle{2}}
\put(113,1021){\line(0,1){6}}
\put(113,1019){\line(0,-1){6}}
\picplus{118}{1031}
\put(118,1032){\line(0,1){6}}
\put(118,1030){\line(0,-1){6}}

\end{picture}
\end{center}

\caption{
Spin averaged $\bbb$ spectrum.  A horizontal line indicates
the experimental energy; experimental values for 
spin-singlet $P$ states are assumed to equal spin average
of the corresponding triplet states.  Results
are shown at $\beta$=$ 5.6$ ($n_f$=$2$ NRQCD), 6.0 ($n_f$=$0$ NRQCD),
6.2 (UK(NR)QCD), and 6.1 (Fermilab).
}
\label{fig:ups_spinav}
\end{figure}

Three groups have calculated more than one splitting 
in the $\bbb$ spectrum 
recently.  
This year, the Fermilab\cite{fnal-spect,khadra-lat94} and 
NRQCD\cite{nrqcd-spect,davies-lat94} groups have increased their quenched
statistics and the NRQCD group has results at $n_f=2$.
The results for spin averaged splittings are shown in \fig{fig:ups_spinav}.  
The \FNAL group determines the lattice spacing 
from the $\oneP$-$\oneS$ splitting; the NRQCD collaboration uses both
the $\oneP$-$\oneS$ and the $\twoS$-$\oneS$ splitting.
The zero of energy in \fig{fig:ups_spinav} is adjusted so that the
$\Ups(\oneS)$ energy matches experiment;
the bare mass of the $b$ quark
is adjusted so that the kinetic mass, $M_2$, of this state agrees with
its experimental value.  


The discretization and relativistic errors in \fig{fig:ups_spinav}
are expected to be about 5\mev for the NRQCD group, 50\mev for 
UK(NR)QCD\cite{uknrqcd},
and somewhere between the two for Fermilab.  This is because the NRQCD
collaboration includes all \order($mv^4$) terms in 
the action, while UK(NR)QCD
only works at \order($mv^2$).  The clover action used by \FNAL has an
incorrect coefficient for the $\pv^4/(8m^3)$ term in the action,
which becomes correct as $am\rightarrow 0$; the error in this
coefficient controls the size of \order($mv^4$) errors in the \FNAL results.

The NRQCD collaboration has spectrum results at $n_f$=$2$.  These 
use an ensemble obtained from the HEMCGC collaboration, with two
flavors of staggered fermions at $am$=$.01$ and 
$\beta$=$5.6$\cite{hemcgc-ens}.  
The \CDHW results, which I will discuss later, used this 
ensemble and a similar one (also from HEMCGC) at $am$=$.025$.  
The only other dynamical results available are from the \KT group,
who worked at $\beta$=$5.7$ rather than $5.6$\cite{kt-ens}.  Both the
\CDHW and \KT groups have results only for the $\oneP$-$\oneS$ splitting.

The HEMCGC collaboration has performed a tremendous public
service by making their ensembles publicly available; given the
high cost of generating dynamical ensemble, 
it is important that as much analysis as possible be done on each
one.  
Many other groups have been similarly generous in providing their
quenched ensembles to the NRQCD collaboration.

The results show excellent agreement with
experiment.  The most important thing to focus on in \fig{fig:ups_spinav}
is the relative size of the $\twoS$-$\oneS$ and $\oneP$-$\oneS$ splittings;
their ratio is plotted in \fig{fig:2s1srat} as a function of $n_f$.
It appears that there is a discrepancy in this 
ratio at $n_f$=$0$, 
$(\twoS$-$\oneS)/(\oneP$-$\oneS)$=$1.41(5)$, when compared with
the experimental value of 1.28, and that this discrepancy is 
removed when extrapolated to an $n_f$ value between $2$ and $4$.  
\fig{fig:2s1srat} also shows 
consistent, but noisier, results at $\beta$=$5.7$.
This is an encouraging indication of scaling, but it should be noted that
this splitting has been adjusted by a large ($\approx$$40\mev$) 
perturbative correction which removes \order($a^2$)
errors in the gluonic propagator\cite{davies-lat94,nrqcd-alphas}; 
The NRQCD collaboration results may also show a quenching effect for 
the $\twoP$-$\oneP$ splitting; this is not supported, 
however, by the UK(NR)QCD results. 



\input spextrap.tex


\fig{fig:2s1srat} exhibits exactly the behavior one 
expects from quenching errors; determining
$\ainv$ from the $\oneP$-$\oneS$ splitting is roughly equivalent to 
matching the force in
the quenched static potential at the Bohr radius to be the same as
the force in full QCD, i.e. $\alpha_{n_f=0}(r_0) $=$ \alpha_{full}(r_0)$
(this is the essence of the \FNAL prescription for correcting quenching
errors in $\alpha_s$ determinations\cite{fnal-alphas}).  
Since the quenched $\beta$-function runs faster than
the full one, the potential $\alpha(r)/r$ will be weaker 
than it should be at short range.  This in turn should shift 
quenched $S$ states up relative to the $P$ states.  
The charm quark is too heavy to affect the running of $\alpha$ at the
Bohr radius of $\Ups$, so the spin-averaged spectrum should agree with 
experiment when extrapolated to $n_f$=$3$.



\begin{figure}[htb]
\begin{center}
\setlength{\unitlength}{.025in}
\begin{picture}(95,95)(15,-60)
\put(15,-50){\line(0,1){80}}
\multiput(13,-40)(0,20){4}{\line(1,0){4}}
\multiput(14,-40)(0,10){7}{\line(1,0){2}}
\put(12,-40){\makebox(0,0)[r]{$-40$}}
\put(12,-20){\makebox(0,0)[r]{$-20$}}
\put(12,0){\makebox(0,0)[r]{$0$}}
\put(12,20){\makebox(0,0)[r]{$20$}}
\put(12,30){\makebox(0,0)[r]{MeV}}
\put(15,-50){\line(1,0){100}}


\put(40,-55){\makebox(0,0)[t]{${^3S}_1$ / ${^1S}_0$}}
\put(70,-55){\makebox(0,0)[t]{${^1P}_1$}}
\put(100,-55){\makebox(0,0)[t]{${^3P}_J$}}

\multiput(28,0)(3,0){7}{\line(1,0){2}}
\put(50,2){\makebox(0,0)[t]{$\Upsilon$}}
\put(35,0){\circle*{2}}
\put(40,0){\circle{2}}
\put(44,-1){\framebox(2,2)}

\put(50,-32){\makebox(0,0)[t]{$\eta_b$}}
\put(35,-39){\circle*{2}}
\put(40,-30){\circle{2}}
\put(44,-26){\framebox(2,2)}

\put(63,-5){\makebox(0,0)[l]{$h_b$}}
\put(70,-3){\circle*{2}}
\put(75,-5){\circle{2}}

\multiput(90,-40)(3,0){7}{\line(1,0){2}}
\put(110,-40){\makebox(0,0)[l]{$\chi_{b0}$}}
\put(97,-33){\circle*{2}}
\put(97,-33){\line(0,1){5}}
\put(97,-33){\line(0,-1){5}}
\put(102,-34){\circle{2}}
\put(102,-33){\line(0,1){6}}
\put(102,-35){\line(0,-1){6}}

\multiput(90,-8)(3,0){7}{\line(1,0){2}}
\put(110,-8){\makebox(0,0)[l]{$\chi_{b1}$}}
\put(97,-12){\circle*{2}}
\put(97,-12){\line(0,1){3}}
\put(97,-12){\line(0,-1){3}}
\put(102,-10){\circle{2}}
\put(102,-9){\line(0,1){3}}
\put(102,-11){\line(0,-1){3}}

\multiput(90,13)(3,0){7}{\line(1,0){2}}
\put(110,13){\makebox(0,0)[l]{$\chi_{b2}$}}
\put(97,14){\circle*{2}}
\put(97,14){\line(0,1){2}}
\put(97,14){\line(0,-1){2}}
\put(102,13){\circle{2}}
\put(102,14){\line(0,1){2}}
\put(102,12){\line(0,-1){2}}

     \put(26, 24){\circle*{2}}
\multiput(26, 20)(0,3){3}{\line(0,1){2}}
     \put(30, 29){\makebox(0,0)[l]{expected}}
     \put(30, 24){\makebox(0,0)[l]{systematic}}
     \put(30, 19){\makebox(0,0)[l]{error}}


\end{picture}
\end{center}
\caption{$\bbb$ spin splittings.  
The expected (non-quenching) systematic error in the 
NRQCD calculation is shown; the \FNAL results should have similar errors.
Symbols are the same as in {\protect\fig{fig:ups_spinav}}. 
}
\label{fig:ups_spinsplit}
\end{figure}


The NRQCD and \FNAL groups have also calculated spin splittings in
the $\bbb$ system; the results are shown in \fig{fig:ups_spinsplit}.
The (non-quenching) systematic errors are expected to be
about $5\mev$ for both groups, and are indicated in the figure.  

The most important feature of \fig{fig:ups_spinsplit} is that the 
overall size of the $\chi_b$ splittings agrees with experiment, within
the expected errors.  
This is significant
for two reasons.  First, both the \FNAL and NRQCD actions are tadpole
improved, which increases the strength of the spin interactions by a 
factor of $u_0^{-3}$; without tadpole improvement the $\chi_b$ splittings are
reduced by 50\%\cite{nrqcd-lat93}.  Second, spin splittings are 
quadratically sensitive 
to the lattice spacing.  This is because they are roughly inversely 
proportional to the quark mass.
Using an incorrectly low value of 
$\ainv$ to compare with experiment will result in tuning to too heavy a 
quark mass; this reduces the dimensionless spin splittings, which are then 
multiplied by an additional factor of (the incorrect) $\ainv$ to make them
dimensionful.  If the correct lattice spacing for $\Ups$ physics at 
$\beta$=$6.0$ were $2.0\gev$ rather than the $2.4\gev$ found by the NRQCD
collaboration, the $\chi_b$ splittings in \fig{fig:ups_spinsplit} would be
roughly 30\% smaller in magnitude, resulting in a 15-20\mev disagreement
with experiment.

The quenched ratio $(\chi_{b2}-\chi_{b1})/(\chi_{b1}-\chi_{b0})$
appears too small compared to experiment; adding dynamical fermions
pushes the ratio the wrong way.  Although this is only a one to two $\sigma$
effect, the same behavior occurs with good statistical significance in 
the $\ccb$ system.
In both $\bbb$ and $\ccb$ the disagreement with experiment is within the 
expected systematic errors.  It will be
interesting to see if improving the NRQCD or \HW actions to
eliminate \order($v^2$) corrections to the spin-splittings will fix this
discrepancy.

Unfortunately, the $\eta_b(1S)$ has not been observed experimentally;
extrapolating the NRQCD results to $n_f$=$3$ (where $\ainv$ and $m_0$ have been
adjusted) results in a hyperfine splitting
of $44{\rm MeV}$\cite{davies-lat94}.  Because the hyperfine splitting is
a short-distance quantity, dynamical charm loops are likely to have an
effect, but a simple perturbative estimate indicates that this is likely
to be at most a few {\rm MeV}.
The result should be somewhere between
$45$ and $50\mev$, which agrees with what is expected
from potential models.


\subsection{$\ccb$}

Both the NRQCD\cite{davies-lat94,nrqcd-charm} and 
\FNAL\cite{khadra-lat94,fnal-spect} groups 
have increased their statistics
for quenched $\ccb$ spectroscopy.  I do not show the spin averaged
spectrum, since the only independent level is the $\twoS$, which
is too noisy to resolve quenching effects.
Instead, the spin-dependent spectrum is shown in
\fig{fig:jpsi_spinsplit}, along with the expected (non-quenching)
systematic errors (estimated to be roughly $35\mev$), which are much
larger than for $\bbb$ since $v^2$ is three times as big.
The interesting thing about \fig{fig:jpsi_spinsplit} is that
these systematic errors are dominant; the statistical errors are negligible.
Within the (rather large) systematic uncertainty, \fig{fig:jpsi_spinsplit}
shows agreement between simulation and experiment.  In addition, the
trends seen in $\bbb$ spin splittings (\fig{fig:ups_spinsplit})
are now quite clear; hyperfine and fine splittings are too small
and the ($\chi_{c2}-\chi_{c1}$)/($\chi_{c1}-\chi_{c0}$) ratio is too big.


\begin{figure}[tb]
\begin{center}
\setlength{\unitlength}{0.012in}
\begin{picture}(190,180)(10,-140)
\put(15,-130){\line(0,1){180}}
\multiput(13,-120)(0,40){5}{\line(1,0){4}}
\multiput(14,-120)(0,10){16}{\line(1,0){2}}
\put(12,-120){\makebox(0,0)[r]{$-120$}}
\put(12,-80){\makebox(0,0)[r]{$-80$}}
\put(12,-40){\makebox(0,0)[r]{$-40$}}
\put(12,0){\makebox(0,0)[r]{$0$}}
\put(12,40){\makebox(0,0)[r]{$40$}}
\put(12,50){\makebox(0,0)[r]{MeV}}

\put(15,-130){\line(1,0){180}}

     \put(170,-40){\circle*{3}}
\multiput(170,-59)(0,3){13}{\line(0,1){2}}
\put(150,-66){\makebox(0,0)[l]{expected}}
\put(150,-76){\makebox(0,0)[l]{systematic}}
\put(150,-84){\makebox(0,0)[l]{error}}

\multiput(28,30)(3,0){5}{\line(1,0){2}}
     \put(57,30){\makebox(0,0)[t]{$J / \Psi$}}
     \put(33,24){\circle{3}}
     \put(35.5,16.3){\framebox(3,3)}

\multiput(28,-89)(3,0){5}{\line(1,0){2}}
     \put(54,-84){\makebox(0,0)[t]{$\eta_c$}}
     \put(33,-72){\circle{3}}
     \put(35.5,-54.75){\framebox(3,3)}
     \put(37,-51.75){\line(0,1){.75}}
     \put(37,-54.75){\line(0,-1){.75}}

\multiput(83,-1)(3,0){5}{\line(1,0){2}}
     \put(90,-9){\circle{3}}
     \put(90,-7.5){\line(0,1){.5}}
     \put(90,-10.5){\line(0,-1){.5}}

\multiput(123,-110)(3,0){5}{\line(1,0){2}}
     \put(128,-75){\circle{3}}
     \put(128,-73.5){\line(0,1){4.5}}
     \put(128,-76.5){\line(0,-1){4.5}}

\multiput(123,-14)(3,0){5}{\line(1,0){2}}
     \put(139,-14){\makebox(0,0)[l]{$\chi_{c1}$}}
     \put(128,-21){\circle{3}}
     \put(128,-19.5){\line(0,1){2.5}}
     \put(128,-22.5){\line(0,-1){2.5}}

\multiput(123,31)(3,0){5}{\line(1,0){2}}
     \put(139,31){\makebox(0,0)[l]{$\chi_{c2}$}}
     \put(130,33){\circle{3}}
     \put(130,34.5){\line(0,1){.5}}
     \put(130,31.5){\line(0,-1){.5}}
\end{picture}
\end{center}
\caption{$\ccb$ spin splittings.  Symbols are same as
{\protect\fig{fig:ups_spinav}}.  Levels are adjusted so
that spin averaged energies agree with experiment.
}
\label{fig:jpsi_spinsplit}
\end{figure}


The bad news from \fig{fig:ups_spinsplit} is that $\ccb$, for which 
there is more experimental data, is not as well controlled a system as
$\bbb$ is.  This was to be expected; in addition to suffering from worse
relativistic corrections, $\ccb$ is a larger meson with softer internal
glue,  and so is more susceptible to finite volume effects
and the particulars of the dynamical fermion content.  
For these reasons, I 
suspect that the $\bbb$ system is probably more reliable for $\alpha_s$
determinations than $\ccb$.

The good news is that $\ccb$ is
a wonderful laboratory for studying improved actions; 
the effects of (currently neglected) \order($mv^6$) and 
\order($\alpha_smv^4$) terms in the NRQCD action are resolvable.  A
drop in the disagreement with experiment of the $\ccb$ spin spectrum 
of a factor of 3 or so (i.e. $v^2$) upon inclusion of these terms (or
similar ones in the \HW approach) would
be a tremendous result showing the efficacy of improved actions.

A final comment: the validity of the \HW formalism at any 
value of $am$ is a definite
advantage for charm physics.  As presently formulated, NRQCD blows up
for charm at roughly (quenched) $\beta$=$5.85$, but has a much more easily
computed improved action.  A \HW action, improved to the some order in
$v^2$ and $\alpha_s$, will almost certainly have 
smaller systematic
errors when run at a high $\beta$ than the similarly improved NRQCD action
run at $\beta_{max} $=$ 5.85$ will have.  On the other hand, the \HW run
(say at $\beta$=$6.4$) will probably cost 100 times as much computer time
at the same physical volume.
Today, with \HW actions in use which are less highly improved 
than the NRQCD actions, the relative size of systematic errors is unclear.
For $\bbb$ physics, this advantage of \HW does not yet enter; even at
$\beta$=$6.4$, $am$$\approx$$1$ and NRQCD can be used.

\section{$M_b$}

Last year, the NRQCD collaboration presented a preliminary result for the pole
mass of the bottom quark\cite{nrqcd-lat93}.  The final results of this 
calculation can be found in \cite{nrqcd-mb}, which also includes the 
effects of dynamical fermions and a determination of
the $\msb$ mass.  This year the \FNAL group has presented
a preliminary determination of the pole mass of the charm quark
\cite{khadra-lat94}.

The basic idea is very simple; given a particular
regularization scheme (i.e. $\msb$, IHW, or NRQCD), the perturbative
pole mass can be calculated in terms of the mass parameter of the theory.
\be
 M = Z_m^S\,\,             m_{0}^S 
        = (1 + C^S\alpha_s(q^*))  m_{0}^S,
\label{pole_mass}
\ee
where $S$ labels the scheme, $C^S$ is calculated in perturbation
theory, and $q^*$ is a scale 
chosen as in \cite{lmac} or \cite{blm}.
The $\msb$ mass can be obtained by equating the pole 
mass in the $\msb$ and lattice schemes;
\be
M^{\msb}(M) = {Z_m^{NR}\over Z_m^{\msb}}\,m_{0}^{NR}
                   = {Z_m^{IHW}\over Z_m^{\msb}}\,m_{0}^{IHW}.
\label{msb_mass}
\ee
The bare mass of the lattice theory (\HW or NRQCD) is determined 
by requiring that the kinetic mass of the ground state match
experiment; the result is substituted into \eq{pole_mass}
to obtain the pole mass or \eq{msb_mass} to obtain the $\msb$ mass.

The NRQCD collaboration obtains a pole mass for bottom of 
$M_b $=$ 5.0(2)$ and an $\msb$ mass of 
$M_b^{\msb}(M_b) $=$ 4.0(1)$\cite{nrqcd-mb}; 
the \FNAL group has a 
preliminary result of $M_c $=$ 1.5(2)$\cite{khadra-lat94}.

A word of caution: the NRQCD collaboration result
has changed since last year.  This is because $q^*$ is actually
much softer than the preliminary estimate used in\cite{nrqcd-lat93};
$\alpha_V(q^*)$$\approx$$.4$ for the bottom quark at $\beta$=$6.0$.  This
soft scale arises because tadpole improvement cancels out the contribution
to $q^*$ 
of much of the hard glue.
A soft $q^*$ also appears in the $\msb$ scheme; applying the
procedure of \cite{blm} to the two-loop $\msb$ expression for the pole 
mass\cite{broadhurst} gives $q^*$$\approx$$.22\,M_b$.
Because of these soft $q^*$s, perturbative uncertainties dominate
the NRQCD determination of $M_b$, and make an NRQCD determination
of $M_c$ impossible.  Soft $q^*$s are less of a problem
in the $\msb$ mass determination; 
the soft glue cancels in the ratio of $Z$s, leaving a $q^*$ closer 
to the cutoff.  It is not clear what
non-perturbative effects there might be in equating the (perturbative)
pole masses of two different schemes, especially since infra-red glue
seems to be playing such an important role.  The \FNAL group has not
found a similar soft $q^*$ in their \HW calculation.

\section{$\alpha_s$}

Last year, El-Khadra gave an excellent review of $\alpha_s$ 
determinations from onium 
spectroscopy\cite{khadra-lat93}, so I will not go into 
detail.  At that meeting, there were quenched results from the NRQCD and 
\FNAL groups, as well as preliminary dynamical results 
from the \KT group, using their $n_f$=$2$ staggered, $\beta$=$5.7$ ensemble.
This year, the KT\cite{kt-alphas}, CDHW\cite{cdhw} and 
NRQCD\cite{nrqcd-alphas} groups have new dynamical results,
using \KT, HEMCGC, and HEMCGC ensembles (previously discussed), respectively.

The necessary steps in a lattice determination of $\amsbar$ are:
\begin{itemize}
\item{1.
Determine $\alpha(q^*)$ in some lattice scheme; in this talk, I will 
use $\alpha_{V,2}(3.41\ainv)$ (defined by the second order expansion
of the plaquette in $\alpha_V$)\cite{nrqcd-alphas}.  
This and other possible choices will be discussed in depth by Michael in his 
talk\cite{michael-lat94}.
}

\item{2.
Determine $\ainv$ by matching some lattice ``measurement'' to experiment.
When using onium spectroscopy, this is usually the 
spin-averaged $\oneP$-$\oneS$ 
splitting.  The NRQCD collaboration has also performed their analysis
using the $\Ups(\twoS)-\Ups(\oneS)$ splitting.  Both splittings
are very insensitive to the bare
quark mass and to spin-dependent discretization errors.
}

\item{3.
Correct $\alpha(q^*)$ to $\alpha^{(n_f)}(q^*)$, where $n_f$ is the number
of quark flavors above threshold {\it at the typical momentum of the quantity
used to set the scale} in step 2.  For example, the gluons holding 
$\ccb$ or $\bbb$ together
have a typical momentum transfer of about $500$ to $1000$ \mev; the
correct physical $n_f$ is 3, even though $q^*_\Box$ is roughly 8\gev at
$\beta$=$6.0$.
}

\item{4.
Convert from $\alpha^{(n_f)}(q^*)$ to $\alpha_{\msb}^{(n_f)}(\mu)$, using
a perturbative matching formula.  This step requires that the $\alpha$ defined
in step 1 above be a good expansion parameter, i.e. $\alpha_{lat}$ is not
satisfactory.
}

\item{5.
Run $\alpha_{\msb}^{(n_f)}(\mu)$ to $\amsbar$, using the 3-loop beta function.
}

\end{itemize}

The steps above separate the various sources of error in the calculation; 
each step is independent of the
others.  In particular, the choice of scheme for $\alpha$ 
(step~1) and the choice of
experimental observable to set the scale (step~2) are totally independent
decisions;  for example the definition of $\alpha$ found in \cite{lsww}
could be combined with a scale determination from $\bbb$ spectroscopy.
A non-perturbative definition of $\alpha$ ($\alpha_{V,2}$ is such a 
definition, $\alpha_V$ determined from the plaquette\cite{lmac} is not) 
can be used in step~1; only in the
conversion to $\msb$ (step~4) is perturbation theory needed.

The situation for quenched determinations of $\amsbar$ from onium has 
not changed 
since last year.  Basically, all groups get results consistent with
$.112(8)$ from $\ccb$ and $.112(5)$ from $\bbb$(see \cite{khadra-lat93} for 
details).  This is because
the dominant error in the calculation comes from step 3, the correction
to physical fermion content using the technique of 
\cite{fnal-alphas}.  The \KT and NRQCD groups have
used this technique to correct their $\nf$=$2$ results to $\nf$=$3$ as well,
obtaining $.112(4)$ (from $\ccb$) and $.115(3)$ (from $\bbb$), respectively.
The agreement between $\ccb$ and $\bbb$ results is non-trivial;
the various determinations disagree before correction
is made for quenching.  

This year, a new method for using dynamical results to correcting quenching 
effects was introduced in\cite{nrqcd-alphas}.
Results from $\nf$=$0$ and
$\nf$=$2$ calculations are run to an arbitrary 
momentum scale near $q^*_\Box$
and $\alpha^{-1}$ is extrapolated linearly to $\nf$=$3$.  This technique
was also used by CDHW.  The two groups obtain results of $.115(2)$ 
(from $\bbb$) and $.108(6)$ (from $\ccb$), respectively.  
The large disagreement in central values is due to a disagreement in the
scale determination by the two groups, which I discuss below.  
The NRQCD collaboration performed separate, independent, 
analyses for the $\oneP - \oneS$ and $\twoS - \oneS$ splittings of $\bbb$;
the central values were almost identical.
The NRQCD collaboration
claims that step 4, the perturbative conversion from $\alpha_{V,2}$ to 
$\alpha_{\msb}$, rather than step 3,
is now the dominant error in the calculation; CDHW, who the Wilson action,
cite both perturbative conversion and scale determination errors as dominant.

Of all the groups, the NRQCD collaboration claims the smallest errors.
There are many potential pitfalls in these $\alpha_s$ calculations; each
of them must be addressed before we can be sure that the uncertainties
are correctly understood.

The first area of concern is in the choice of scheme for $\alpha$.  In order
for the calculation to be reliable, the conversion from $\alpha$ to 
$\alpha_{\msb}$ must be accurate, i.e. $\alpha$ must be a good perturbative
parameter.  In addition, the observable chosen to determine $\ainv$ must 
exhibit {\it asymptotic} scaling in $\alpha$;
the ratio $aE/(a\Lambda_\alpha)$, where $E$ is the observable 
and $\Lambda_\alpha$ is
the $\Lambda$ parameter of the $\alpha$ scheme, must have negligible 
dependence on $\beta$.  
This is obvious; if it were not the case then
determinations of $\alpha$ at different $\beta$ using the same observable
would yield different values of $\amsbar$.
Michael will compare various prescriptions for $\alpha$ in his talk; 
I discuss asymptotic scaling in the next section.

A second area of concern is that the dynamical fermion content could
be incorrect.  The HEMCGC configurations used by the NRQCD and \CDHW 
groups had fermion bare masses of $am$=$.01$ 
and $am$=$.025$, i.e. about 25 and 60 
{\rm MeV}, while the \KT ensemble at $am$=$.01$ corresponds to about 30 \mev
dynamical quarks.  The NRQCD collaboration has assumed that the high
momentum transfer inside the $\Ups$ ($.5$-$1.0${\rm MeV}) will cause these
these non-physical bare mass values to have negligible effect;
this should be checked explicitly.  In $\ccb$, where the
momentum transfers are not as high, \CDHW see a slight dependence on
the sea quark mass; obtaining $\ainv$'s of  $1.90(9)$ and $1.72(10)$
at $.01$ and $.025$, respectively\cite{cdhw}.  
There are also the usual worries about HMD and the `square root trick'
for ensembles with two staggered flavors.  It is important that other
calculations be done, both with two Wilson and four staggered flavors.
A four staggered flavor calculation would also allow a check of the 
linear extrapolation of $\alpha^{-1}$ in $n_f$.

The third major source of concern, and the most troubling, is the
discrepancy between lattice spacings obtained from NRQCD simulations
of $\bbb$ and those obtained by other methods.  

Last year, the NRQCD
collaboration reported that $\bbb$ spectroscopy gave $\ainv$$\approx$$2.4$
\gev at quenched $\beta$=$6.0$, significantly 
larger than typical light spectroscopy values of $1.8$ - $2.1$ {\rm GeV}.
This is not alarming since the quenched spectrum should {\it not}
agree with the real world; the $\beta$ function runs too quickly so
$\ainv$ from a high energy observable (like $\bbb$ spectroscopy)
should be larger than $\ainv$ from a low energy observable (like
$\ccb$ or $m_{\rho}$).

What is disturbing is that dynamical results seem to exhibit the same
range of $\ainv$'s as quenched. The HEMCGC collaboration found a very good 
match between light spectroscopy on
their $\beta$=$5.6$, $n_f$=$2$ staggered ensemble and a quenched ensemble
at $\beta$$\approx$$5.95$\cite{hemcgc-match};
the NRQCD collaboration obtained a $\bbb$ $\ainv$ of about $2.4$ \gev
on the dynamical ensemble, the same as at quenched $\beta $=$ 6.0$.  One would
naively expect a much smaller spread of momenta at $n_f$=$2$ if quenching
were the culprit; perhaps $2.2$-$2.3$ \gev when $\ainv_{\bbb} $=$ 2.4$ \gev.

My own (very biased) opinion is that the problem probably lies in the light
spectroscopy calculations.  This is mainly because we know that systematic
errors are much more difficult to control for light quarks than for heavy;
\order($a\lqcd$) scaling violations arising from the Wilson action are
one obvious candidate.  NRQCD, on the other hand, appears to have systematic
errors under control.  The spectroscopy results discussed above
showed no big surprises, and the fact that the spin splittings appear to be
correct is important, since they depend quadratically on the ratio of the
calculated lattice spacing to its true value.
The final reason for my bias is that the $n_f$ extrapolation of NRQCD
results is consistent with experiment.  
\fig{fig:2s1srat} implies that the $\oneP$-$\oneS$ and
$\twoS$-$\oneS$ splittings give different $\ainv$ at $n_f$=$0$ but agree
at $n_f$=$3$.  This is the first spectroscopic calculation I'm aware
of in which quenching effects are observed and then removed when
dynamical fermions are turned on.

There are three additional calculations which I think 
need to be done before we can be
confident that the lattice determination of $\amsbar$ from onium
really is within $2\%$ of the true value.  First, the NRQCD calculation
using $\bbb$ spectroscopy
should be repeated at $am$=$.025$; this will
explicitly check the dependence on sea quark mass. 
Second, a similar calculation must be done using $\ccb$ 
spectroscopy.  This will have to be done at lower $\beta$ since NRQCD
blows up for charm on the HEMCGC ensemble.
If $\ccb$ spectroscopy yields the same value of $\alpha$ at physical $n_f$
this will be a beautiful example of the removal
of quenching effects (since $\ainv$s from quenched $\bbb$ and $\ccb$
differ by about $15$\%).  Third, $\bbb$ and $\ccb$ calculations must also 
be done at $n_f$=$2$ 
using the \HW formulation {\it at the same order as the NRQCD calculation}.  
\HW spectroscopy using the tadpole-improved 
\SW action\cite{khadra-lat94} 
tends to give $\ainv$'s about $10\%$ smaller than NRQCD, both for $\bbb$
and $\ccb$; it is important to know whether this is because the two
actions in use have different levels of improvement or if it really
is a disagreement.  At the moment, the news here is bad; the $\ainv$
obtained from $\ccb$ by \CDHW (with no clover term) on the HEMCGC 
configurations is about the same as quenched results.  It
will be interesting to see if a clover \HW run on these configurations
changes this behavior.  El-Khadra has estimated the various corrections
here by using a potential model\cite{khadra-lat94}.

It is important to stress that this lattice spacing problem is not 
just of concern in $\alpha_s$ calculations.  For example, 
both heavy and light quarks appear in
calculations of $f_B$, discussed by Sommer\cite{sommer-lat94}.
$f_B\sqrt{M_B}$ scales like $a^{-3/2}$; changing $\ainv$ from 
$1.8$ to $2.4$\gev corresponds to a 40\% uncertainty.
It is very important that we resolve this issue.

\section{Asymptotic Scaling}

As discussed above, a particular observable must exhibit asymptotic
scaling before it is useful in an $\alpha_s$ determination.  
The $V$ scheme is particularly convenient for studying asymptotic
scaling, since 
$\alv$ is
easily determined from the plaquette\cite{lmac}.  Another advantage
of expressing lattice energies in terms of $\alv$ is the ease of 
comparison between groups; it would be very difficult to find highly
accurate determinations of $m_\rho$, for example, at the various 
$\beta$ values I discuss below; this is not a problem for the plaquette.


\begin{figure}[htb]
\centerline{\ewxy{}{60mm}}
\caption{
Asymptotic scaling plot for the quenched $\bbb$ spectrum.
Flat dependence (up to logs) on $(\alv)^2$ indicates asymptotic scaling; 
linear dependence indicates \order($a^2$) discretization errors.
Simulation value of $\twoS$-$\oneS$ splitting has been rescaled by the
experimental ratio ($\oneP$-$\oneS$/$\twoS$-$\oneS$).
A perturbative correction has been applied to remove the effects of 
\order($a^2$) errors in the gluonic action.  Points at $\beta$=$5.7$ have
been spread horizontally for clarity.
}
\label{fig:gla_ups_sp_corr}
\end{figure}


The dependence of the spin-averaged quenched $\bbb$ splittings on 
$(\alv)^2$ is shown in \fig{fig:gla_ups_sp_corr}.  The vertical axis is
\be
{\Delta {\overline M}\over \Lambda_V} 
= \left({(a\Delta M)_{sim}\over \alv}\right)  
\left({452\mev\over (\Delta M)_{exp}}\right),
\label{deltm_def}
\ee
where $(a\Delta M)_{sim}$ is the lattice value of the splitting being
plotted, $(\Delta M)_{exp}$ is the experimental value of that splitting,
and 452\mev is the experimental value of the $\bbb$ $\oneP$-$\oneS$
splitting (assuming a hyperfine splitting of $50\mev$).  The rescaling
by experimental values
is equivalent to comparing dimensionless ratios to experiment;
if two splittings have the same value of $a\Delta {\overline M}$, then 
their ratio matches experiment and they will give the same value of $\ainv$.
I have chosen $(\alv)^2$ as the horizontal axis because the leading
discretization errors in both the FNAL and NRQCD collaboration results
should be \order($a^2$).
All values were adjusted to correct for \order($a^2$)
errors in the gluonic action.  All lattices have physical volume much
larger than $\bbb$ mesons.

There are three different sets of splittings in \fig{fig:gla_ups_sp_corr}
(\HW and NRQCD $\oneP$-$\oneS$ and NRQCD $\twoS$-$\oneS$);
all exhibit asymptotic scaling within errors.  The only
point that is inconsistent with flat behavior is the \HW point at $\beta$=$5.7$;
both NRQCD splittings at this $\beta$ are consistent with $6.0$ (albeit
with large error bars for $\twoS$-$\oneS$).  This
may be because the NRQCD collaboration removes the dominant \order($a^2$)
errors in their action ($C_5$ and $C_6$);  see \cite{khadra-lat94} for
an estimate of the effects of these terms on the \FNAL results.

Although the results in \fig{fig:gla_ups_sp_corr} exhibit asymptotic scaling,
they do not agree on $\ainv$!  The discrepancy between $\oneP$-$\oneS$ 
and $\twoS$-$\oneS$ within NRQCD is good news; this is the quenching
effect discussed in \fig{fig:2s1srat}.  The disagreement between NRQCD
and \HW for the $\oneP$-$\oneS$ splitting will be bad news if
it doesn't disappear when the \HW calculation is repeated with all 
\order($mv^4$) improvements turned on. 
If it doesn't go away and if the scale in the NRQCD  
$\alpha_s$ calculation is 10\% too high at $n_f$=$3$, then
the value of $\amsbar$ in \cite{nrqcd-alphas} is shifted down by 2\%.


\begin{figure}[htb]
\centerline{\ewxy{}{60mm}}
\caption{
Same as {\protect\fig{fig:gla_ups_sp_corr}}, except the
perturbative correction has not been applied.
}
\label{fig:gla_ups_sp}
\end{figure}


At low $\beta$, there is an important caveat to the above discussion.  All 
of the points in \fig{fig:gla_ups_sp_corr} had a perturbative correction
applied to remove the effects of \order($a^2$) errors in the gluonic 
ensembles\cite{nrqcd-alphas}; the uncorrected results are shown 
in \fig{fig:gla_ups_sp}.
The main effect is to move the $\beta$=$5.7$ $\oneP$-$\oneS$ points down 
significantly; the $\beta$=$5.9$ \FNAL point is also shifted down 
about two $\sigma$.
The $\twoS$-$\oneS$ splitting is negligibly affected, as are $\ccb$ 
splittings.

It is obviously dangerous to rely too heavily on points
where this correction is large when drawing conclusions.  I have 
chosen to apply it in these scaling plots, however, because we know that
it is there and a similar calculation of the hyperfine splitting works
well\cite{nrqcd-alphas}.  It is important to obtain direct results using
an improved gluonic ensemble; Lepage showed in his talk that the one-loop
improved gluonic action is tremendously successful at removing \order($a^2$) 
errors\cite{lepage-lat94}.  It will be much cheaper to check asymptotic
scaling in $\bbb$ by using an improved gluonic action at 
$\beta$=$5.7$ than by trying to get high statistics at $\beta$=$6.4$.


\begin{figure}[htb]
\centerline{\ewxy{}{60mm}}
\caption{
Same as {\protect\fig{fig:gla_ups_sp_corr}}, except displays
quenched $\ccb$ spectrum.  All splittings shown are $\oneP$-$\oneS$, and
have been rescaled by experimental ratio between $\bbb$ to $\ccb$.
Gluonic correction has been applied.  NRQCD point at $\beta$=$5.7$ has been 
moved horizontally for clarity.
}
\label{fig:gla_psi_spect_corr}
\end{figure}


Results for the $\oneP$-$\oneS$ splitting in $\ccb$ are presented in 
\fig{fig:gla_psi_spect_corr}; $\twoS$-$\oneS$ results are too noisy
to be of interest.  
The discrepancy between NG and non-NG splittings
in the HS formulation is obvious; the two results bracket the \HW value.
There is only one $\beta$ with NRQCD results; NRQCD should blow up at
$(\alv)^2 < .05$.  The \HW results seem to be scaling better than for
$\bbb$; see \cite{khadra-lat94} for a discussion.
The high $\beta$ \HW splittings are roughly 10\% higher than the low $\beta$
NRQCD splitting.  I have no idea if this is related to 
the similar behavior in the
$\bbb$ system; \order($mv^6$) terms left out of the NRQCD action should
contribute at the 10\% level for $\ccb$.
It would be very interesting to
go to the next order of improvement for both the NRQCD and \HW actions in
this system.

Note that $\Delta{\overline M} / \Lambda_V$ is roughly 15\% larger for
$\ccb$ than for $\bbb$.  Much of this is probably due to quenching.
Systematic errors in the $\ccb$ splittings could also be noticeably 
contributing, however.

I combine $\ccb$ and $\bbb$ $n_f$=$2$ results in 
\fig{fig:gla_nf2_corr}.  Two things are striking.  First, the NRQCD
$\bbb$ splittings are in much closer agreement than in 
\fig{fig:gla_ups_sp_corr}.  Again, this the same information 
as \fig{fig:2s1srat}, and indicates quenching errors are being removed.
Second, the NRQCD $\bbb$
results and \CDHW $\ccb$ results disagree by roughly the same amount
as in Figs.~\ref{fig:gla_ups_sp_corr} and \ref{fig:gla_psi_spect_corr}.
This is the reason for the disagreement between the two groups' central
values of $\amsbar$.  Unfortunately, the \KT group has much larger statistical
errors.  Both the \CDHW and \KT groups used the Wilson, rather
than \SW action; it would be interesting to see the \CDHW run repeated using
a clover, asymmetric clover, or \order($mv^4$) \HW action. 


\begin{figure}[htb]
\centerline{\ewxy{}{60mm}}
\caption{
Same as {\protect\fig{fig:gla_ups_sp_corr}}, except displays
$n_f$=$2$ spectrum.  NRQCD points are $\bbb$, all others are $\ccb$.  
Splittings have been rescaled by experimental ratio.
NRQCD $\oneP$-$\oneS$ 
and \KT staggered points have been moved horizontally for clarity. \KT
staggered points are a lower bound, due to finite volume effects.
}
\label{fig:gla_nf2_corr}
\end{figure}



\section{Conclusions}

We are well on the way to having an excellent understanding of
HH mesons.  We already have many cross-checks; more should be
forthcoming.

I wish to thank my colleagues in the Fermilab, Kyoto-Tsukuba,
\CDHW and NRQCD collaborations for their patience and work 
in supplying me 
with and explaining their results.  This talk was prepared at the
University of Glasgow, during a visit supported by the PPARC.
I am extremely grateful to the members of the Physics Department 
there, especially Christine Davies, for their hospitality during
my stay.  This work was supported in part by a DOE grant.

\input bib.tex
\end{document}

%% file: spextrap.tex
\begin{figure}[thb] 
\begin{center}
\setlength{\unitlength}{.015in}
\begin{picture}(170,120)(20,0)
\put(35,0){\line(0,1){91}}
\multiput(33,0)(0,30){4}{\line(1,0){4}}
\multiput(34,0)(0,15){6}{\line(1,0){2}}
\put(32,0){\makebox(0,0)[r]{1.0}}
\put(32,30){\makebox(0,0)[r]{1.2}}
\put(32,60){\makebox(0,0)[r]{1.4}}
\put(32,90){\makebox(0,0)[r]{1.6}}
\put(35,0){\line(1,0){140}}
\multiput(55,-2)(30,0){4}{\line(0,1){4}}
\put(57,-7){\makebox(0,0)[r]{0}}
\put(87,-7){\makebox(0,0)[r]{1}}
\put(117,-7){\makebox(0,0)[r]{2}}
\put(147,-7){\makebox(0,0)[r]{3}}
\put(177,-9){\makebox(0,0)[r]{{\bf $n_f$}}}
\put(50,105){\makebox(0,0)[r]{$ {\bf \frac {2S-1S} {1P-1S}}$}}
\multiput(15,42)(5,0){30}{\line(1,0){2}}
\multiput(145,0)(0,5){18}{\line(0,1){2}}
\put(55,61){\circle*{4}}
\put(115,49){\circle*{4}}
\put(55,61){\line(0,1){8}}
\put(55,61){\line(0,-1){8}}
\put(115,49){\line(0,1){8}}
\put(115,49){\line(0,-1){8}}
\put(55,61){\line(5,-1){90}}
\put(50,60){\circle{4}}
\put(50,62){\line(0,1){13}}
\put(50,58){\line(0,-1){13}}

\end{picture}
\end{center}

 \caption{Dependence on $n_f$ of the ratio of
the $2S-1S$ to $1P-1S$ splittings in $b\overline{b}$.
The filled circles give values at quenched $\beta$ = 6.0
and $\beta$=5.6, $n_f$=2.  
The open circle is quenched 
$\beta$ =5.7 for comparison. 
The horizontal dashed line
gives the experimental value of 1.28 and solid line is an extrapolation to
$n_f$ = 3, drawn to guide the eye.
All results are from the
NRQCD collaboration{\protect\cite{davies-lat94}}. 
}
\label{fig:2s1srat}
\end{figure}

%% file: heavyq94.bbl
\begin{thebibliography}{9}

\bibitem{cornell-action} G.P. Lepage et al., 
    { Phys. Rev.} D46 (1992) 4052 and references therein.

\bibitem{fnal-action} A.X. El-Khadra, A.S. Kronfeld and P.B. Mackenzie,
in preparation.

\bibitem{lepage-lat91} G.P.~Lepage,
   Nucl. Phys. B (Proc. Suppl.) 26 (1992) 45

\bibitem{mackenzie-lat92} P.B.~Mackenzie
   Nucl. Phys. B (Proc. Suppl.) 30 (1993) 35

\bibitem{davies-lat93} C.T.H.~Davies,
   Nucl. Phys. B (Proc. Suppl.) 34 (1994) 135

\bibitem{lmac} G.P. Lepage and P.B. Mackenzie,  {Phys. Rev.} D48 (1993) 2250.

\bibitem{kronfeld-lat92} A.S.~Kronfeld, 
   Nucl. Phys. B (Proc. Suppl.) 30 (1993) 445

\bibitem{sw-action} B.~Sheikholeslami and R.~Wohlert,
   Nucl. Phys. B 259 (1985) 572

\bibitem{khadra-lat94} A.X.~El-Khadra, these proceedings.

\bibitem{uknrqcd}   S.M.~Catterall et al., Phys. Lett. 321B (1994) 246.

\bibitem{sinclair-lat934} G.T.~Bodwin, S.~Kim and D.K.~Sinclair,
   Nucl. Phys. B (Proc. Suppl.) 34 (1994) 434; these proceedings.

\bibitem{kt-alphas} S.~Aoki et al., UTHEP-280 (1994), hep-lat-9407015.

\bibitem{onogi-lat93} T.~Onogi et al.,
   Nucl. Phys. B (Proc. Suppl.) 34 (1994) 492

\bibitem{cdhw} M.~Wingate et al., these proceedings.

\bibitem{davies-lat94} NRQCD Collaboration: C.T.H.~Davies et al., 
these proceedings.

\bibitem{nrqcd-alphas} C.T.H.~Davies et al., OHSTPY-HEP-T-94-013
(1994) hep-ph-9408328, to appear in Phys. Lett. B.

\bibitem{nrqcd-mb}   C.T.H.~Davies et al., { Phys. Rev. Lett.} 73 (1994) 2654.

\bibitem{kronfeld-lat93} A.S.~Kronfeld and B.P.~Mertens, 
   Nucl. Phys. B (Proc. Suppl.) 34 (1994) 495

\bibitem{kronfeld-lat94} A.S.~Kronfeld, these proceedings.

\bibitem{fnal-loop} A.X.~El-Khadra, A.S.~Kronfeld, P.B.~Mackenzie
and B.~Mertens, in progress.

\bibitem{morning-notad} C.~Morningstar,
   Nucl. Phys. B (Proc. Suppl.) 34 (1994) 425; { Phys. Rev.} D48 (1993) 2265.

\bibitem{morning-qstar}   C.~Morningstar, { Phys. Rev.} D50 (1994) 5902.

\bibitem{mertens-priv}   B.~Mertens, private communication.

\bibitem{aoki-lat94} S.~Aoki et al., these proceedings.

\bibitem{kt-ens} M.~Fukugita et al.,
                 { Phys. Rev.} D47 (1993) 4739.

\bibitem{fnal-spect} A.X.~El-Khadra et al., in progress.

\bibitem{nrqcd-spect} C.T.H.~Davies et al., OHSTPY-HEP-T-94-005
(1994) hep-lat-9406017, to appear in Phys. Rev. D.

\bibitem{hemcgc-ens} K.~Bitar et al.,
   Nucl. Phys. B (Proc. Suppl.) 26 (1992) 259; {Phys. Rev.} D46 (1992) 2169.


\bibitem{fnal-alphas}   A.X.~El-Khadra et al., 
   { Phys. Rev. Lett.} 69 (1992) 729.

\bibitem{nrqcd-lat93} NRQCD Collaboration: G.P.~Lepage et al.,
   Nucl. Phys. B (Proc. Suppl.) 34 (1994) 417

\bibitem{nrqcd-charm} C.T.H.~Davies et al., in preparation.

\bibitem{blm} S.J.~Brodsky, G.P. Lepage and P.B.~Mackenzie, 
                 { Phys. Rev.} D28 (1983) 228.

\bibitem{broadhurst} N.~Gray et al., Z.~Phys. {\bf C48}, 673 (1990). 

\bibitem{khadra-lat93} A.X.~El-Khadra
   Nucl. Phys. B (Proc. Suppl.) 34 (1994) 449

\bibitem{michael-lat94} C.~Michael, plenary talk, these proceedings.

\bibitem{lsww} M. L\"uscher et al.,Nucl.Phys. B359(1991)221;
Nucl. Phys. B413 (1994) 481.

\bibitem{hemcgc-match} K.M.~Bitar et al., 
                 { Phys. Rev.} D48 (1993) 370.

\bibitem{sommer-lat94} R.~Sommer, plenary talk, these proceedings.

\bibitem{lepage-lat94} G.P.~Lepage, these proceedings.


\end{thebibliography}
